\documentclass[journal=nalefd,manuscript=article,layout=twocolumn]{achemso}

\usepackage[version=3]{mhchem} 
\usepackage[T1]{fontenc}
\usepackage{subcaption} 
\usepackage{textcomp} 
\usepackage{amsmath}
\usepackage{booktabs} 

\newcommand{\FontSizeMain}{\footnotesize}
\newcommand{\FontSizeCaption}{footnotesize}

\author{Andrew Malouf}
\email{andrew.malouf@adelaide.edu.au}
\author{Ori Henderson-Sapir}
\affiliation[University of Adelaide]
{Department of Physics and Institute of Photonics \& Advanced Sensing, University of Adelaide, Adelaide}
\author{Sze Yun Set}
\author{Shinji Yamashita}
\affiliation[University of Tokyo]
{Research Center for Advanced Science and Technology, University of Tokyo, Tokyo}
\author{David J. Ottaway}
\affiliation[University of Adelaide]
{Department of Physics and Institute of Photonics \& Advanced Sensing, University of Adelaide, Adelaide}

\title{Two-Photon Absorption and Saturable Absorption of Mid-IR in Graphene}

\abbreviations{IR,mid-IR,near-IR,CVD,SA,2PA,PMMA,TLS,FTIR}

\begin{document}
\FontSizeMain
\begin{abstract}
We report on the response of graphene to high intensity mid-IR radiation and show that graphene exhibits saturable absorption and significant two-photon absorption in the spectral region from 1.55\,\textmu m to 3.50\,\textmu m (0.35\,eV to 0.80\,eV). We find that the effective modulation depth of multilayer graphene is limited by two-photon absorption which will affect its performance as a laser mode-locking element. The measured saturation intensities of femtosecond pulses were found to depend on the third power of photon energy when we combined our results with others reported in literature, while those of longer pulses were found to have a square root dependence.
\end{abstract}



\section{Introduction}

The optical and electronic properties of graphene are important for a number of applications including solar cells, photodetectors, light-emitting devices, and touch screens \cite{bonaccorso2010graphene}. Graphene has also been used extensively as a saturable absorber to mode-lock ultrafast lasers because it has broadband saturable absorption (SA), large modulation depth, and ultrafast relaxation time \cite{bao2009atomic,tan2010mode,sun2010graphene,martinez2011mechanical,ugolotti2012graphene,cizmeciyan2013graphene,yamashita2014short,zhu2016graphene}. Graphene mode-locked fiber lasers were first demonstrated in the telecommunication band \(  \sim  \)1.5\,\textmu m (0.8\,eV) \cite{bao2009atomic,sun2010graphene,martinez2011mechanical} and subpicosecond pulses were achieved. The longest wavelength demonstrated in a graphene mode-locked laser is 2.8\,\textmu m (0.4\,eV) in which 42\,ps pulses were achieved  \cite{zhu2016graphene}. There is currently an urgent need to understand the mid-IR saturation properties of graphene to facilitate growing demand for mode-locked mid-IR laser sources in fields such as molecular spectroscopy and medicine \cite{vainio2016mid,serebryakov2010medical}.

Graphene has a variety of interesting electronic properties including linear dispersion near the Dirac points, zero bandgap, and charge carriers that have speed \(  v\approx10^{6}\,\mathrm{ms^{-1}}  \) and mimic relativistic particles with zero rest mass \cite{novoselov2005two}. The electronic band structure can be described using a tight-binding Hamiltonian \cite{wallace1947band}. Pauli blocking at high intensity combined with ultrafast response times and linear dispersion make graphene an ideal broadband saturable absorber for passive mode-locking \cite{bonaccorso2010graphene,vasko2010saturation,marini2017theory}. Compared to traditional semiconductor saturable absorber mirrors (SESAMs) and single-walled carbon nanotubes, graphene has an extremely broad wavelength response, high modulation depth and low saturation intensity \cite{bao2009atomic}. The modulation depth can be extended by stacking multiple graphene layers \cite{sobon2015multilayer,zhang2015dependence}.

The saturation intensity, \(  I_{\mathrm{s}}  \), has been suggested to have a wavelength \(  \lambda  \) dependence given by the empirical relationship \(  {I_{\mathrm{s}}=2.7/\lambda^{6}}  \), where \(  I_{\mathrm{s}}  \) and \(  \lambda  \) are expressed in units of \(  \mathrm{GW/cm^{2}}  \) and \(  \mathrm{\text{\textmu} m}  \) respectively \cite{yamashita2014short}. The empirical fit was made to saturation intensities measured using wavelengths between 780\,nm and 1560\,nm (0.8\,eV to 1.6\,eV). However, this relationship did not hold for some extremely small reported values of \(  I_{\mathrm{s}}  \) \cite{bao2009atomic,tan2010mode}. We present a revised empirical fit to \(  I_{\mathrm{s}}  \) measured with wavelengths between 435\,nm and 3.5\,\textmu m (0.4\,eV to 2.9\,eV).

Two-photon absorption (2PA) has been observed in the spectral region between 435\,nm and 1100\,nm (1.1\,eV to 2.9\,eV) \cite{yang2011giant,chen2015two}. The strength of 2PA is greater in stacked layers than single layer graphene due to the increased number of energy bands caused by interlayer coupling and thus a greater number of possible electronic transitions \cite{yang2011giant}. The effective modulation depth of graphene is limited by 2PA which can be detrimental to passive mode-locking \cite{grange2005new}.

In this work, we study the saturation behavior of trilayer graphene mounted on \ce{CaF2} using the well known z-scan technique to measure intensity dependent transmission \cite{sheik-bahae1990sensitive,chapple1997single}. We use a 100\,fs tunable laser source at a range of wavelengths, ranging from 1.55\,\textmu m (0.80\,eV), where measured saturation intensity is compared with values in literature \cite{bao2009atomic,martinez2011mechanical,yamashita2014short}, up to 3.5\,\textmu m (0.35\,eV), where recent advances have been made in fiber lasers \cite{henderson-sapir2017recent,henderson-sapir2016versatile,fortin2016watt,jobin2018gain,yang2018cascade,qin2018black,malouf2016numerical}. We show that 2PA limits the effective modulation depth, as well as the slope of the nonlinear transmission curve, at mid-IR wavelengths. We also show that saturation intensities of femtosecond pulses follow the empirical relation \(  {I_{\mathrm{s}}\propto E_{\mathrm{ph}}^3}  \) where \(  E_{\mathrm{ph}}  \) is the photon energy, while for longer pulses, \(  {I_{\mathrm{s}}\propto \sqrt{E_{\mathrm{ph}}}}  \). To the best of our knowledge, this is the first reported measurement of the transmission of high-intensity radiation through graphene using a single sample and a fixed pulse duration over a wide spectral region in the near to mid-IR.

The SA and 2PA processes are described by Eqn. \ref{Eqn:dIdzSA2PA} where \( I=I\left(z'\right) \) is the intensity and \( z' \) is the depth into the material. The absorption parameters \( \alpha_{0} \), \( \alpha_{\mathrm{ns}} \), and \( \beta \) are the saturable, non-saturable, and 2PA parameters respectively. The non-saturable coefficient \(  \alpha_{\mathrm{ns}}  \) is included because saturable absorbers are generally imperfect and do not saturate absorption down to zero \cite{boyd2008nonlinear,keller1996semiconductor}. 
\begin{equation}
\frac{\mathrm{d}I}{\mathrm{d}z'}=-\left[\frac{\alpha_{0}}{1+\frac{I}{I_{\mathrm{s}}}}+\alpha_{\mathrm{ns}}+\beta I\right]I
\label{Eqn:dIdzSA2PA}
\end{equation}
We make no assumptions about the change in intensity \textit{within} the sample and treat the trilayer thickness as dimensionless and unity such that \(  \mathrm{d}z'=1  \). The transmission \( T \) is then described by Eqn. \ref{Eqn:T_SA2PA},
\begin{equation}
T\left(I\right)=\exp\left[-\left(\frac{\alpha_{0}}{1+\frac{I}{I_{\mathrm{s}}}}+\alpha_{\mathrm{ns}}+\beta I\right)\right]T_\mathrm{sub}
\label{Eqn:T_SA2PA}
\end{equation}
where \( T_{\mathrm{sub}} \) is the transmission through the substrate. Note that in this expression, the units of \(  \beta  \) are inverse intensity and the SA parameters \(  \alpha_{0}  \) and \(  \alpha_{\mathrm{ns}}  \) are dimensionless.

\section{Experiment}
The intensity dependent transmission of mid-IR radiation through graphene was measured using the z-scan technique. Graphenea (Spain) produced the graphene using chemical vapor deposition (CVD) and transferred three monolayers separately onto the face of a 25\,mm diameter, 5\,mm thick \ce{CaF2} window (Thorlabs WG51050). The tunable light source was an optical parametric amplifier (Light Conversion TOPAS-C) pumped by an 800\,nm regenerative amplifier system (Spectra Physics Spitfire Pro XP). The pulse duration was 100\,fs full width at half maximum and the repetition rate was 1\,kHz. The intensity dependent transmission was measured at six wavelengths - 1550\,nm, 2000\,nm, 2500\,nm, 2800\,nm, 3200\,nm, and 3500\,nm.

\begin{figure}
	\includegraphics[width=8cm]{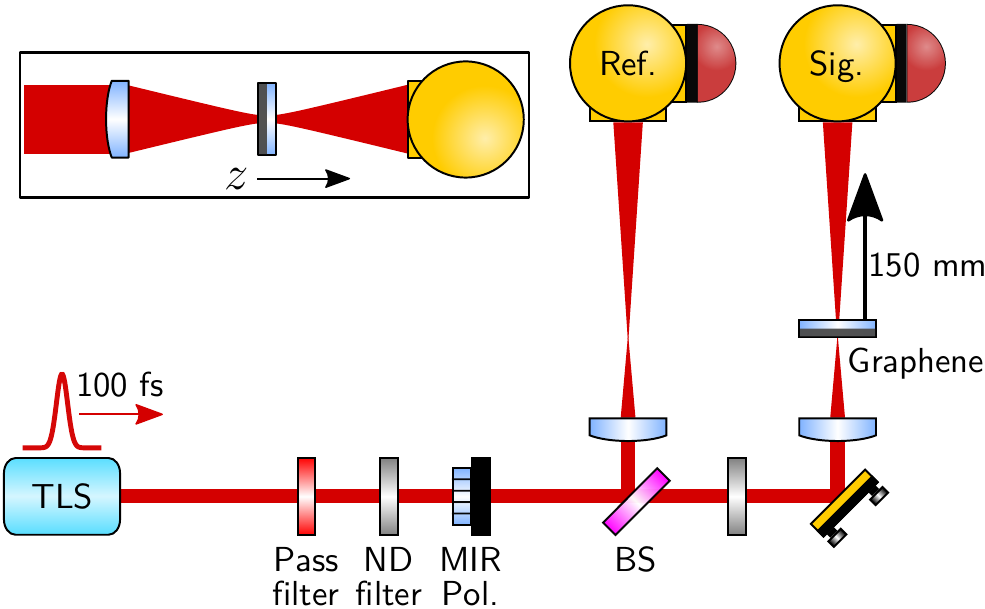}%
	\caption{Schematic of z-scan experiment. The graphene sample was translated along \(  z  \) from the beam waist by up to 150\,mm. The filtered beams were attenuated by neutral density (ND) filters to below the damage threshold of graphene. The mid-IR (MIR) polarizer prevented polarization drifts incident on the beamsplitter (BS).}%
	\label{Fig:Schematic}
\end{figure}

The experiment setup is illustrated in Figure \ref{Fig:Schematic}. The beam was split into a reference path and signal path using a beamsplitter (BS, Thorlabs BSW511). Each beam was focussed using a \ce{CaF2} plano convex lens (\(  f{=}75  \)\,mm). The spot radius at the focal point was between 65\,\textmu m and 78\,\textmu m depending on the wavelength. Each detection system included a gold integrating sphere (Newport 819D-GL-4) and PbSe detector (Thorlabs PDA20H-C), suitable for the 1.5\,\textmu m to 4.8\,\textmu m spectral region. The integrating spheres were implemented to minimize the effects of beam jitter. The beams that entered each integrating sphere were matched in size and closely matched in power to ensure comparable detector response.

At each sample position, \(  z  \), the pulse energies of the signal and reference were measured simultaneously, both with and without the sample in place. The sample position was controlled by a motorized translation stage and motorized flip mount.
In this way, the time between measurements was minimized, reducing effects from fluctuations in the source power. See Supporting Information for more detail on the z-scan procedure.

\begin{figure}
	\includegraphics[width=8cm]{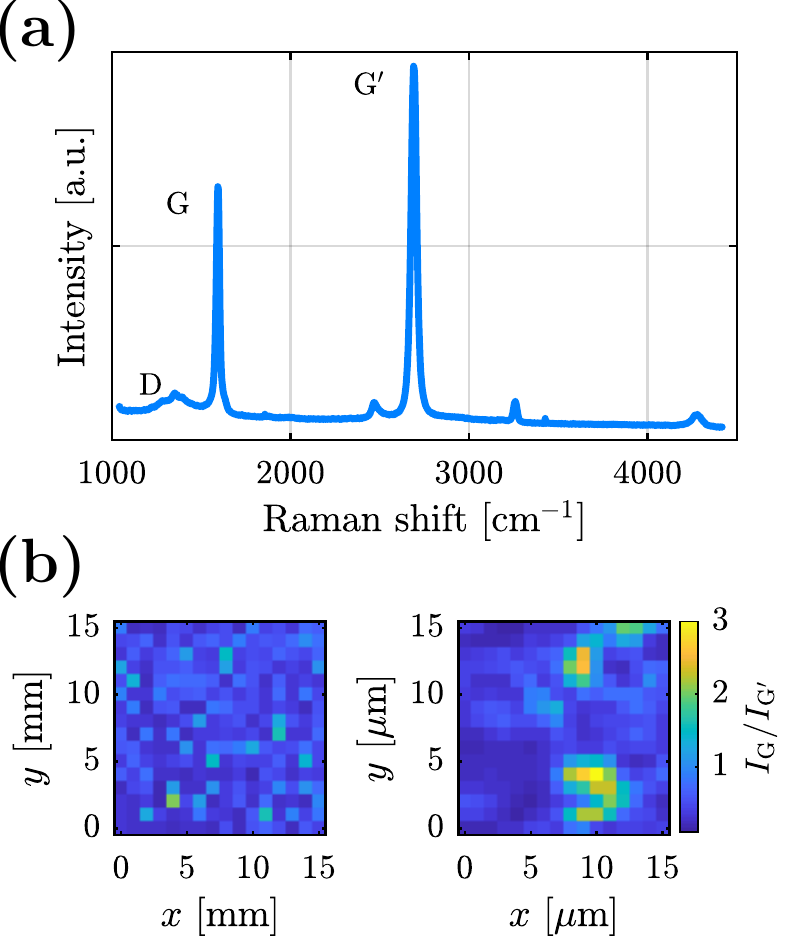}
	\caption{(a) Raman spectra of single layer CVD graphene used to produce the trilayer sample used in this experiment. The graphene was transferred onto a \ce{SiO2}/\ce{Si} substrate. The spectrum was taken with a laser excitation wavelength of 532\,nm (2.33\,eV). (b) Raman maps of the central areas of the graphene sample at \(  1\,\mathrm{mm}  \) (left) and \(  1\,\mathrm{\text{\textmu} m}  \) (right) spacings taken with a laser excitation wavelength of 532\,nm (2.33\,eV). The \(  1\,\mathrm{mm}  \) spacing map covers a large portion of the sample area with low spatial resolution, while the \(  1\,\mathrm{\text{\textmu} m}  \) spacing map covers the central region with high spatial resolution. The color scale shows the ratio \(  I_{\mathrm{G}}/I_{\mathrm{G'}}  \) of integrated peaks.}%
	\label{Fig:Raman}
\end{figure}

The quality of the graphene sample was assessed using Raman spectroscopy and scanning Raman microscopy. The Raman spectra of the CVD single layer graphene used to produce the trilayer sample, as provided by the supplier (Graphenea, Spain), are presented in Figure \ref{Fig:Raman}a. These single layers were stacked on a \ce{CaF2} substrate to form the trilayer graphene used in this experiment. The most prominent features in the Raman spectra of graphene are the G band near \(  \mathrm{1600\,cm^{-1}}  \) and G\('\) band near \(  \mathrm{2700\,cm^{-1}}  \). The large intensity of the G\('\) band relative to the G band is explained by a triple resonance process that can occur due to graphene's linear dispersion \cite{malard2009raman}. The D band arises from the breathing modes of the hexagonal \(  \mathrm{sp^2}  \) carbon rings and requires the presence of a defect for its activation \cite{pollard2014quantitative}.

Raman maps were measured at 1\,mm and 1\,\textmu m spacings after the experiment was completed to analyze the degrees of uniformity and disorder in the graphene sample. The laser excitation wavelength used for the Raman maps was 532\,nm (2.33\,eV) with a \(  {\sim}1\,\mathrm{\text{\textmu}m}  \) spot size. The ratio of the integrated G band to G\('\) band over the area of the sample is shown by the color map in Figure \ref{Fig:Raman}b. The variations may be explained by local changes in stacking orientation which affect the degree of coupling between layers \cite{kim2012raman} and variations in distance between layers. Although the quality of the graphene is high, it is not crystalline. The orientation of each layer is random and an inhomogeneous residue of polymethyl methacrylate (PMMA), resulting from the wet-transfer of CVD graphene, may exist between the layers. The higher ratio of \(  I_{\mathrm{G}}/I_{\mathrm{G'}}  \) exhibits the Raman signature of a strong coupling between the layers while the lower ratio resembles single layer graphene. The  ratio \(  I_{\mathrm{D}}/I_{\mathrm{G}}\approx0.1  \) was averaged over all Raman map locations indicating that no significant disorder was introduced since the graphene was produced \cite{pollard2014quantitative}.

The Raman maps show that large variations in the degree of coupling between graphene layers occur on the scale of several \(  \mathrm{\text{\textmu}m^2}  \). As the smallest spot radius used in the z-scan transmission measurements was \(  65\,\mathrm{\text{\textmu}m}  \), any effects from variations in layer coupling are averaged over the relatively large beam area and are thus not likely to change with beam location. See Supporting Information for Fourier transform infrared (FTIR) spectra of the graphene sample.

\section{Results and Discussion}

\begin{figure}
	\includegraphics[width=1.0\columnwidth]{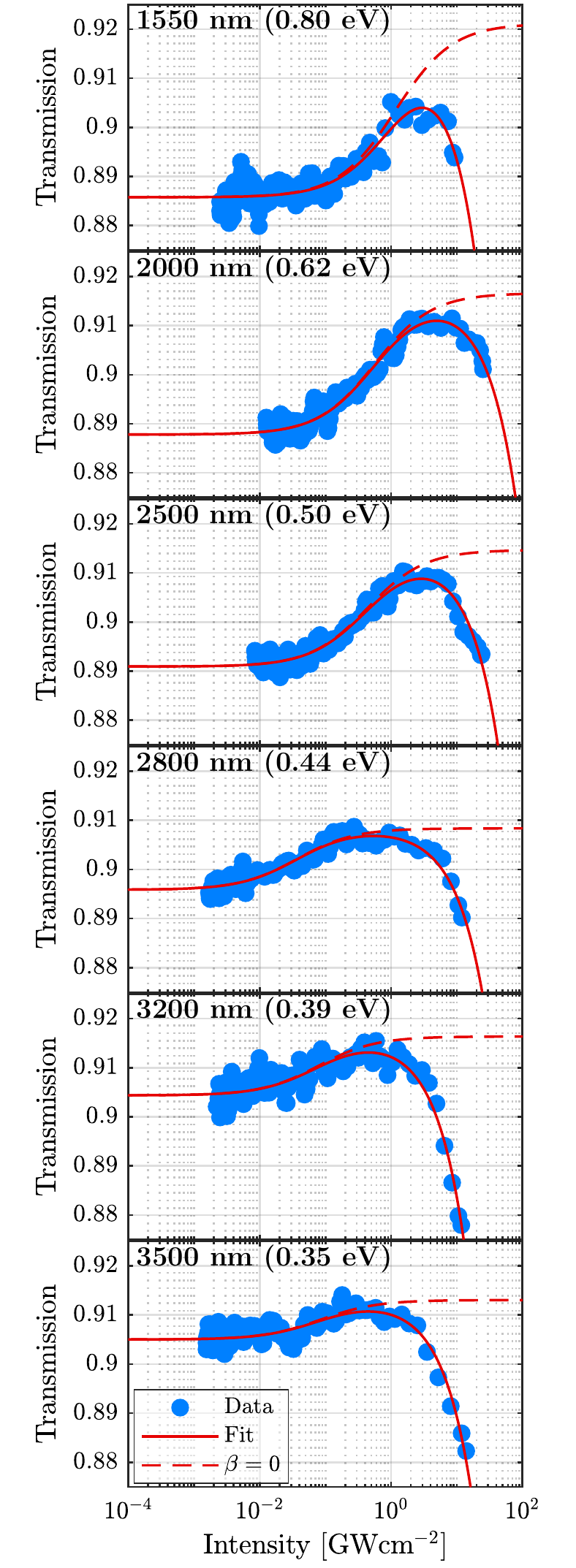}%
	\caption{Z-scan transmission measurements presented as functions of intensity. The data is fitted to Eqn. \ref{Eqn:T_SA2PA} and the absorption processes SA and 2PA are observed. The fit function is then altered by setting \(  \beta=0  \) to qualitatively show the effect of SA without 2PA.}%
	\label{Fig:Z-scan-T_vs_I}
\end{figure}

The transmission \(  T  \) of femtosecond pulses at each wavelength were measured at each position \(  z  \) and converted to a function of intensity \(  I  \). The transmission data were then fitted to Eqn. \ref{Eqn:T_SA2PA} and are presented in Figure \ref{Fig:Z-scan-T_vs_I}. The data at all wavelengths show an increase in transmission with intensity that is consistent with SA. At intensities above \(  \mathrm{{\sim}1\,GWcm^{-2}}  \), the \(  I^2  \) dependence of 2PA dominates the effects of SA and the transmission rolls off. At wavelengths 2.8\,\textmu m (0.44\,eV), 3.2\,\textmu m (0.39\,eV), and 3.5\,\textmu m (0.35\,eV), the transmission reduces to well below unsaturated values. Similar observations have been made in the near-IR regime with bilayer graphene \cite{yang2011giant} as well as SESAMs \cite{grange2005new}. The highest peak intensities incident on the graphene were limited to below damage thresholds determined by experiment at each wavelength. The femtosecond laser induced damage threshold tests were performed on a sacrificial sample of single layer graphene mounted on \ce{CaF2}. See Supporting Information for listed damage thresholds.

\begin{figure}
	\includegraphics[width=1\columnwidth]{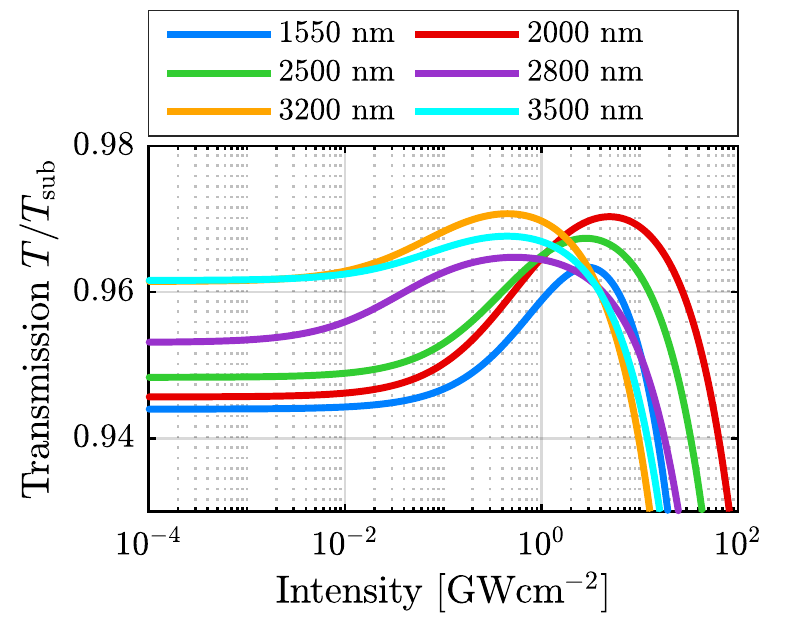}%
	\caption{Transmission fit functions normalized to the transmission of the \ce{CaF2} substrate at each wavelength.}%
	\label{Fig:SA2PA-fit-functions-normalised}
\end{figure}

The transmission functions fitted to Eqn. \ref{Eqn:T_SA2PA} were normalized to the transmission of the \ce{CaF2} substrate at each wavelength and are presented in Figure \ref{Fig:SA2PA-fit-functions-normalised}. There are several interesting features displayed here. Firstly, the low intensity transmission increases with wavelength, which is also observed in the FTIR spectra presented in the Supporting Information. This is in agreement with some reports in literature \cite{cizmeciyan2013graphene,hu2017large}, however it does contradict other reports of a completely flat absorption spectrum \cite{nair2008fine,dawlaty2008measurement} and may be due to an interaction between the graphene and the \ce{CaF2} substrate. Secondly, the saturation intensity decreases with wavelength with the exception of 2.8\,\textmu m (0.44\,eV) where \(  I_{\mathrm{s}}  \) is the lowest of all wavelengths. Thirdly, the modulation depth is highest at 2.0\,\textmu m (0.62\,eV) and lowest at 3.5\,\textmu m (0.35\,eV). The effective modulation depth and slope of the nonlinear transmission curve are reduced by 2PA.

\begin{figure}
	\includegraphics[width=1\columnwidth]{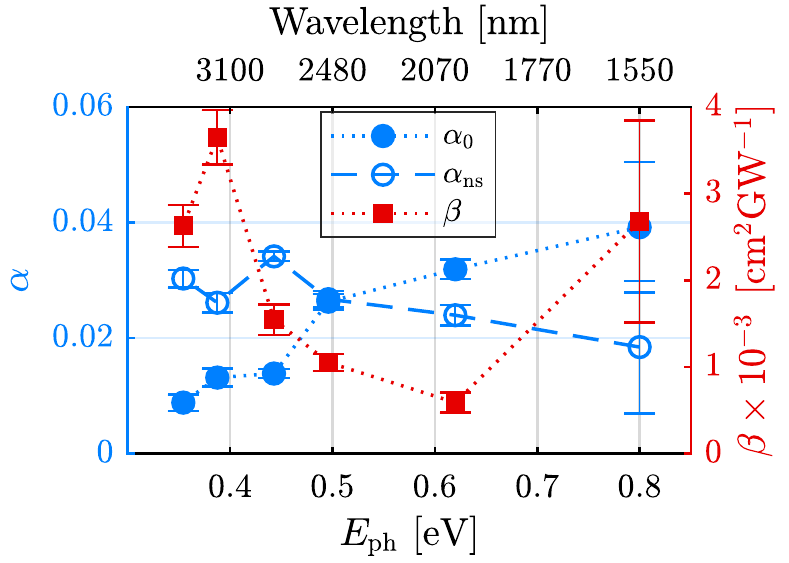}%
	\caption{Fit parameters \(  \alpha_{\mathrm{0}}  \) and \(  \alpha_{\mathrm{ns}}  \) for SA (left axis) and \( \beta  \) for 2PA (right axis). Connecting lines are included for a guide to the eye.}%
	\label{Fig:SA&2PA-fit-params-Eph}
\end{figure}

The fitted absorption parameters for SA and 2PA are shown graphically in Figure \ref{Fig:SA&2PA-fit-params-Eph}. Resonant features in 2PA are observed with a peak at around 3200\,nm (0.39\,eV) which may be explained by interlayer coupling. This peak location agrees with quantum perturbation theory used for the case of AB stacked bilayer graphene \cite{yang2011giant}. There is insufficient data to resolve a possible second peak below 2000\,nm (above 0.62\,eV) that may exist due to three layer coupling. The sum of the SA parameters decreases with wavelength which corresponds to the transmission increase in the low intensity regime. See Supporting Information for listed values of the fitted absorption parameters \(  \alpha_{\mathrm{0}}  \), \(  \alpha_{\mathrm{ns}}  \), and \(  \beta  \).

\begin{figure}
	\includegraphics[width=1\columnwidth]{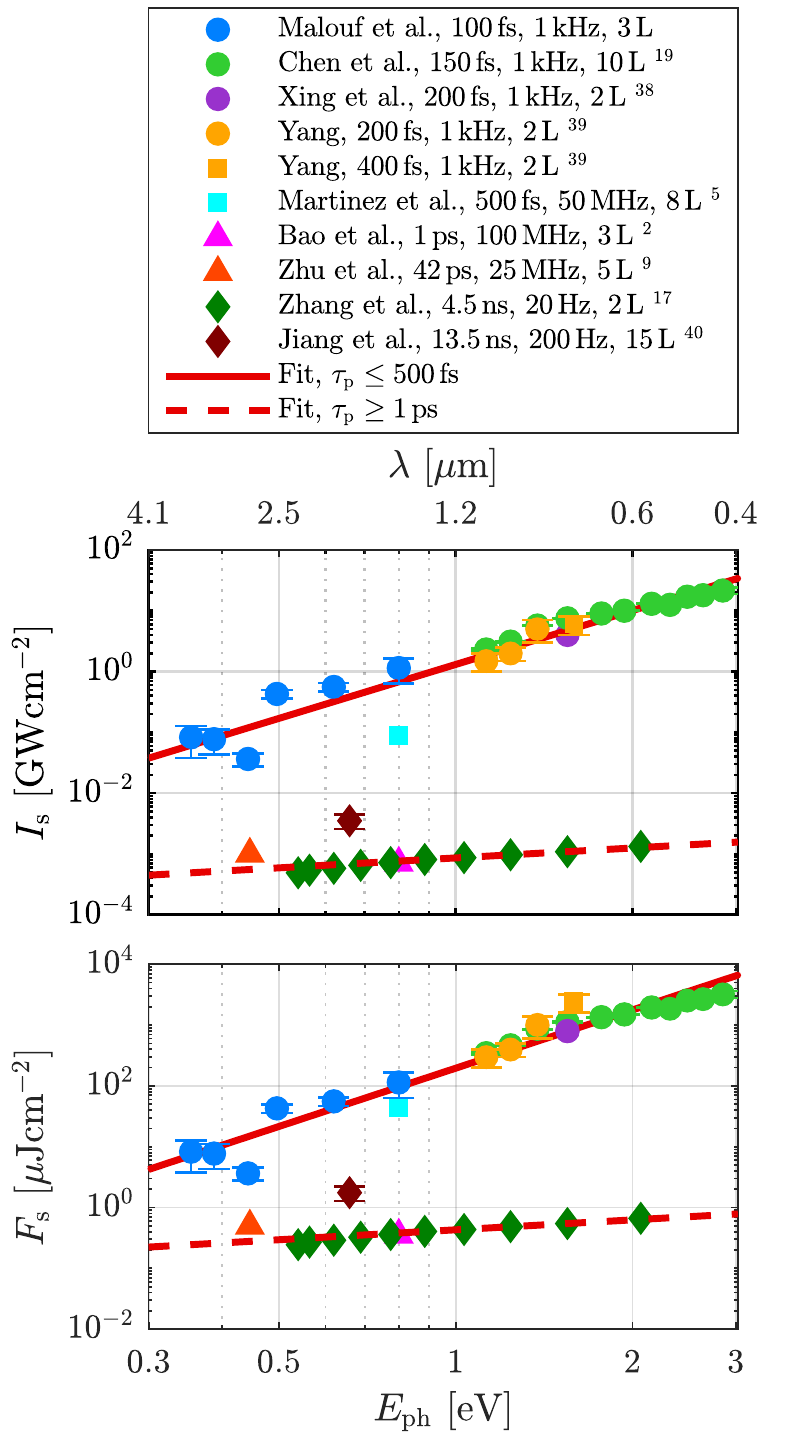}%
	\caption{Comparison of measured saturation intensities \(  I_{\mathrm{s}}  \) and saturation fluences \(  F_{\mathrm{s}}  \) in literature as functions of photon energy \(  E_{\mathrm{ph}}  \). The data are sorted by laser pulse duration. Laser pulse repetition rates and number of graphene layers (L) are listed in the legend. The data are divided into two pulse duration regimes, \(  \tau_{\mathrm{p}}\leq500\,\mathrm{fs}  \) and \(  \tau_{\mathrm{p}}\geq1\,\mathrm{ps}  \), and fitted separately.}
	\label{Fig:Saturation-intensity-fluence}
\end{figure}

The measured saturation intensities, \(  I_{\mathrm{s}}  \), and saturation fluences, \(  F_{\mathrm{s}}  \), are presented in Figure \ref{Fig:Saturation-intensity-fluence} as functions of photon energy, \(  E_{\mathrm{ph}}  \) (See Supporting Information for the listed values), and compared with measurements published in literature \cite{chen2015two,xing2010physics,yang2011giant,yang2012saturable,martinez2011mechanical,bao2009atomic,zhu2016graphene,zhang2015dependence}. In each case, graphene sheets were produced by either mechanical exfoliation, epitaxy, or CVD and the direction of incident radiation was perpendicular to the graphene plane. Cases for graphene flakes suspended in a solution or polymer were excluded from the analysis because the angle of incidence is not uniform and the incident beam is more likely to interact with flake edges where disorder is high. One reported measurement of \(  I_{\mathrm{s}}  \) for graphene produced by spin coating is also included in Figure \ref{Fig:Saturation-intensity-fluence} for comparison, since the graphene was mounted on a flat glass substrate and the radiation was at normal incidence \cite{jiang2018low}.

The relationship between \(  I_{\mathrm{s}}  \) and \(  E_{\mathrm{ph}}  \) is affected by incident pulse duration, \(  \tau_{\mathrm{p}}  \), relative to the carrier lifetime. The relaxation of photogenerated carriers in graphene is described by two distinct time scales, \(  {\tau_{1}\approx140\,\mathrm{fs}}  \)  and \(  {\tau_{2}\approx1.56\,\mathrm{ps}}  \), where \(  \tau_{1}  \) may be attributed to carrier-carrier intraband scattering while \(  \tau_{2}  \) may be explained by carrier-phonon intraband scattering or electron-hole recombination \cite{shang2012probing,dawlaty2008measurement,bao2009atomic}. The fractional amplitudes of the biexponential fits are \(  A_{1}\approx0.74  \) for time constant \(  \tau_{1}  \) and \(  A_{2}  \) for \(  \tau_{2}  \) \cite{shang2012probing}, which result in a mean lifetime of \(  {\tau_{\mathrm{avg}}\approx500\,\mathrm{fs}}  \). The saturation intensity data in Figure \ref{Fig:Saturation-intensity-fluence} are divided into two pulse duration regimes, \(  {\tau_{\mathrm{p}}\leq500\,\mathrm{fs}}  \) and \(  {\tau_{\mathrm{p}}\geq1\,\mathrm{ps}}  \), where the boundary between the two regimes is comparable with the mean carrier lifetime. Each set of data were fitted to \(  {I_{\mathrm{s}}=aE_{\mathrm{ph}}^{b}}  \) and \(  {F_{\mathrm{s}}=cE_{\mathrm{ph}}^{d}}  \), where \(  a,b,c \) and \(  d  \) are the fit parameters. The saturation intensity of graphene produced by spin coating \cite{jiang2018low} is significantly higher than the case for CVD graphene \cite{zhang2015dependence} when measured with similar photon energy and pulse duration, which is likely due to high disorder as a result of the spin-coating process. Therefore, \(  I_{\mathrm{s}}  \) and \(  F_{\mathrm{s}}  \) for the spin-coated graphene were excluded from the fits.

In the short pulse regime, the \(  I_{\mathrm{s}}  \) parameters are \(  {a_{1}=1.3\pm0.2}  \) and \(  {b_{1}=2.9\pm0.2}  \) while the \(  {F_{\mathrm{s}}=\tau_{\mathrm{p}}I_{\mathrm{s}}}  \) parameters are \(  {c_{1}=200\pm22}  \) and \(  {d_{1}=3.2\pm0.2}  \). In the long pulse regime, \(  {F_{\mathrm{s}}=\tau_{\mathrm{avg}}I_{\mathrm{s}}}  \), and the fit parameters are \(  {a_{2}=8.6\pm0.5\times10^{-4}}  \), \(  {b_{2}=0.54\pm0.10}  \), \(  {c_{2}=0.43\pm0.02}  \), and \(  {d_{2}=0.54\pm0.10}  \). That is, the empirical fits suggest that \(  {I_{\mathrm{s}}\propto E_{\mathrm{ph}}^{3}}  \) and \(  {F_{\mathrm{s}}\propto E_{\mathrm{ph}}^{3}}  \) when incident pulse durations are below the mean carrier lifetime. In the case of longer pulses, \(  {I_{\mathrm{s}}\propto\sqrt{E_{\mathrm{ph}}}}  \) and \(  {F_{\mathrm{s}}\propto\sqrt{E_{\mathrm{ph}}}}  \).

\section{Conclusion}
We have characterized the response of trilayer graphene to high intensity radiation between 1.55\,\textmu m and 3.50\,\textmu m (from 0.35\,eV to 0.80\,eV). We have shown that multilayer graphene exhibits SA and 2PA in response to 100\,fs pulses. Resonant features in 2PA were observed over the spectral region measured, however more data is required to resolve these features. The 2PA limits the effective modulation depth and can be detrimental to mode-locking ultrafast lasers. Saturation intensities of femtosecond pulses are shown empirically to be proportional to the third power of photon energy, while those of longer pulses are shown to have a square root dependence.
\\
\\
\begin{acknowledgement}
\FontSizeMain
\\
The authors thank Tak W. Kee and Patrick Tapping for the femtosecond laser facilities and Jason Gascooke for performing Raman measurements. The authors are grateful to Elizaveta Klantsataya for useful discussions. The authors acknowledge the expertise, equipment, and support provided by the Australian National Fabrication Facility (ANFF) at Flinders University. This research was supported by the Australian Research Council through ARC LIEF Grant LE098974 and the South Australian Government Premier’s Research and Industry Fund (PRIF).
\end{acknowledgement}

\begin{suppinfo}
Z-scan procedure, transmission measurements, absorption and saturation intensity values, method of beam profile measurement, damage thresholds, and FTIR spectra.
\end{suppinfo}

\bibliography{2PA_SA_Graphene}

\begin{tocentry}
	\centering
	\includegraphics[]{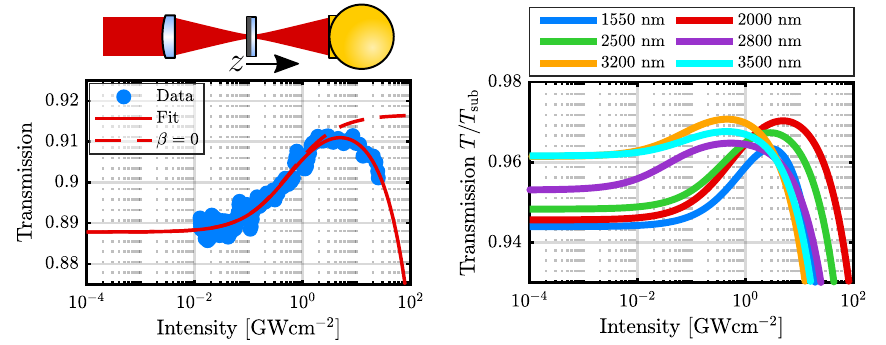}
\end{tocentry}

\end{document}


\section{Z-scan procedure}
The sample size, lens focal length, and z-scan range were chosen to produce an intensity range that spanned 4 orders of magnitude. The beams were filtered to remove any residual pump and idler components. A mid-IR polarizer (Thorlabs LPMIR050) prevented the possibility of polarization drifts proceeding to the beamsplitter and varying the ratio of reflectance to transmittance.

The graphene sample was initially placed at the focal point of the signal lens and translated up to 150\,mm in the direction of beam propagation, \(  \vec{z}  \). The sample was translated from \(  z=0\,\mathrm{mm}  \) to \(  z=50\,\mathrm{mm}  \) in \(  0.5\,\mathrm{mm}  \) increments and from \(  z=50\,\mathrm{mm}  \) to \(  z=150\,\mathrm{mm}  \) in \(  1.0\,\mathrm{mm}  \) increments. A 50\,mm motorized translation stage (Thorlabs MTS50/M-Z8) was used to increment the sample along \(  \vec{z}  \) from each of three fixed positions that extended the range to 150\,mm in total.

The sample was held by a motorized flipper (Thorlabs MFF101/M) that was mounted on the translation stage and enabled the sample to be flipped into and out of the signal beam. Customized software developed in MATLAB was used to command the motorized stage, motorized flipper, and an oscilloscope (Rigol DS4012). The average of 2048 waveforms were acquired from the reference and signal channels at each position \(  z  \) and each flipper position. Transmission \(  T^{(z)}  \) at each position \(  z  \) was calculated from the expression
\begin{equation}
T^{(z)}=\frac
{E_{\mathrm{sig}}^{(z)}/E_{\mathrm{ref}}^{(z)}}
{E_{\mathrm{sig0}}^{(z)}/E_{\mathrm{ref0}}^{(z)}}
\end{equation}
where \(  E_{\mathrm{sig}}^{(z)}  \) and \(  E_{\mathrm{ref}}^{(z)}  \) are the averaged pulse energies detected by the signal and reference respectively at position \(  z  \) when the sample is in the signal beam, and \(  E_{\mathrm{sig0}}^{(z)}  \) and \(  E_{\mathrm{ref0}}^{(z)}  \) are the energies detected when the sample is removed from the beam path. Note that although four pulse energies were measured at each position \(  z  \), only \(  E_{\mathrm{sig}}^{(z)}  \) has any dependence on \(  I  \) (and therefore \(  z  \)).

\section{Transmission measurements}
\begin{figure}
\centering
\includegraphics[]{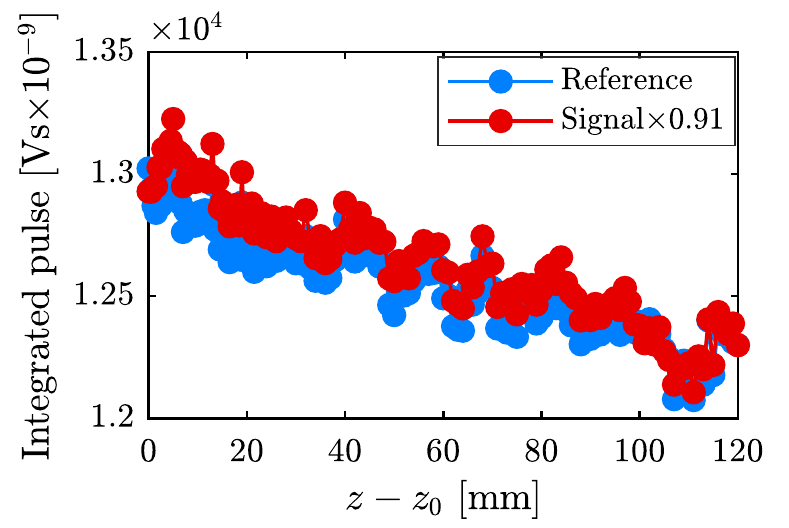}%
\caption{Integrated pulses of the 2800\,nm TLG z-scan.}
\label{Fig:2800IntegratedPulses}%
\end{figure}

\begin{figure}
\centering
\includegraphics[width=1.0\columnwidth]{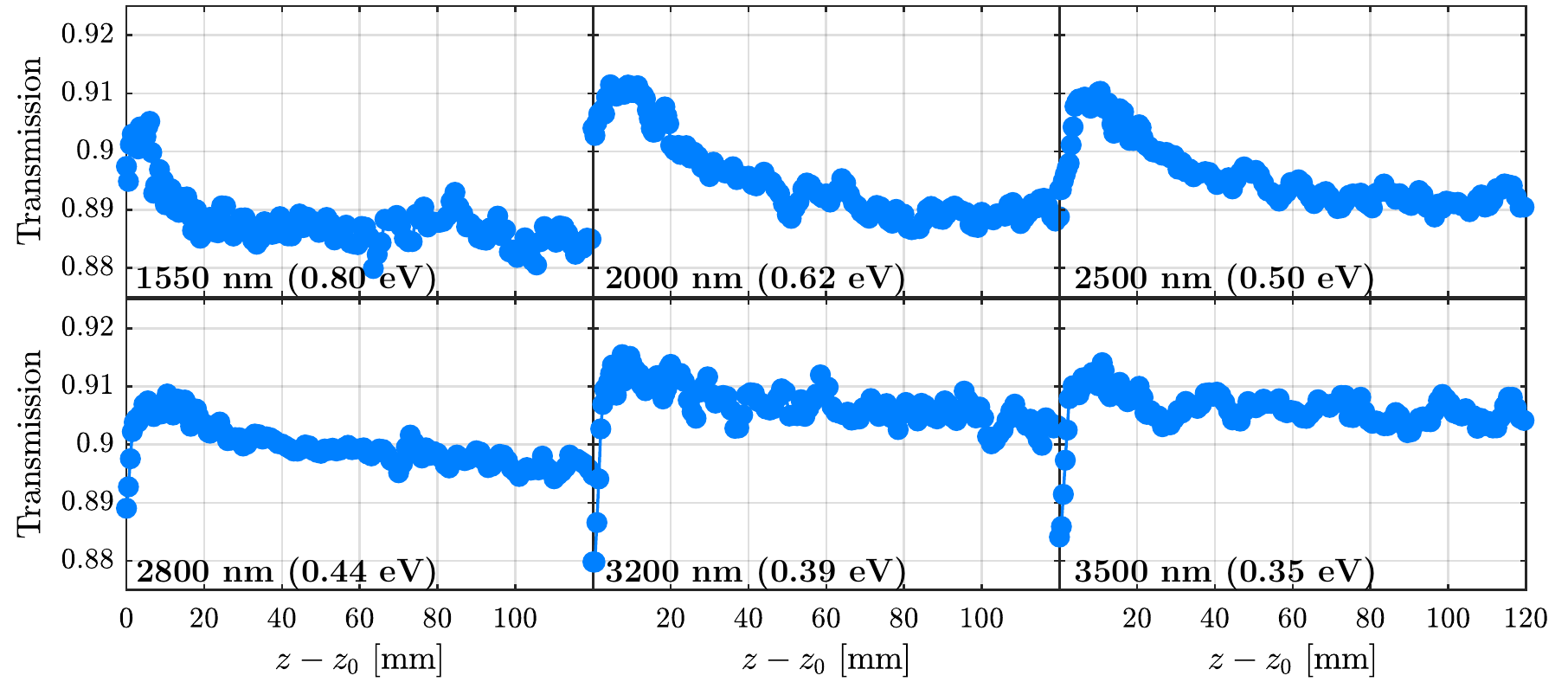}%
\caption{Z-scan transmission data as a function of position \(  z  \). The highest intensity of each scan is at \(  z-z_{0}=0  \).}%
\label{Fig:Z-scan-T_vs_z_2x3}
\end{figure}

The pulse energies were calculated by integrating the averaged waveforms acquired from the oscilloscope. The integrated pulses for the 2800\,nm graphene transmission measurement is presented in Fig. \ref{Fig:2800IntegratedPulses} as an example. The transmission data for each wavelength, calculated from integrated pulses, is presented in Fig. \ref{Fig:Z-scan-T_vs_z_2x3}.

\begin{figure}
\centering
\includegraphics[]{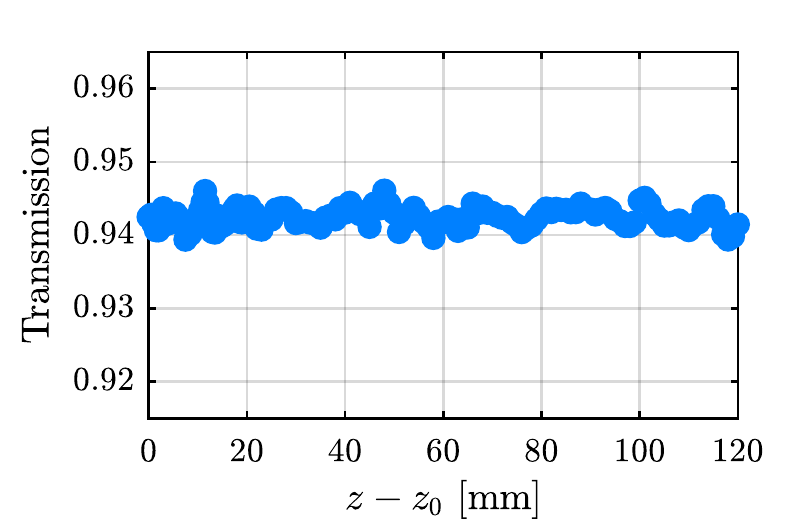}%
\caption{Transmission of 2800\,nm through a \ce{CaF2} window.}
\label{Fig:2800TCaF2}%
\end{figure}

The transmission through a \ce{CaF2} window (Thorlabs WG51050) was also measured at each wavelength to rule out any intensity dependence of the substrate. No intensity dependence was observed. The transmission through the \ce{CaF2} window at 2800\,nm is presented in Fig. \ref{Fig:2800TCaF2} as an example.

\section{Absorption parameters}
The SA parameters \(  \alpha_{\mathrm{0}}  \) and \(  \alpha_{\mathrm{ns}}  \) and 2PA parameter \(  \beta  \) that were fitted to the z-scan transmission measurements are listed in Table \ref{Tbl:FittedAbsorptionParameters}.

\begin{table}
	\caption{Fitted absorption parameters.}
	\label{Tbl:FittedAbsorptionParameters}
	\FontSizeTable
	\begin{tabular}{cccc}
		\toprule
		\begin{tabular}[c]{@{}c@{}}\(\lambda\)\\ \relax[nm]\end{tabular} & \(\alpha_0\times10^{-2}\) & \(\alpha_{\mathrm{ns}}\times10^{-2}\) & \begin{tabular}[c]{@{}c@{}}\(\beta\times10^{-3}\)\\ \relax [cm\textsuperscript{2}GW\textsuperscript{-1}]\end{tabular} \\
		\midrule
		1550 & \(3.92\pm1.13\) & \(1.85\pm1.15\) & \(2.68\pm1.16\) \\
		2000 & \(3.19\pm0.17\) & \(2.40\pm0.18\) & \(0.59\pm0.17\) \\
		2500 & \(2.63\pm0.14\) & \(2.67\pm0.14\) & \(1.05\pm0.10\) \\
		2800 & \(1.39\pm0.08\) & \(3.41\pm0.08\) & \(1.55\pm0.18\) \\
		3200 & \(1.32\pm0.16\) & \(2.61\pm0.17\) & \(3.65\pm0.32\) \\
		3500 & \(0.89\pm0.14\) & \(3.03\pm0.15\) & \(2.63\pm0.24\) \\
		\bottomrule
	\end{tabular}
\end{table}

\section{Saturation intensities}
The saturation intensities \(  I_{\mathrm{s}}  \) that were fitted to the z-scan transmission data are listed in Table \ref{Tbl:FittedSaturationIntensity} along with calculated saturation fluences \(  F_{\mathrm{s}}  \).

\begin{table}
	\caption{Fitted saturation intensities.}
	\label{Tbl:FittedSaturationIntensity}
	\FontSizeTable
	\begin{tabular}{ccc}
		\toprule
		\(\lambda\) & \(I_{\mathrm{s}}\) & \(F_{\mathrm{s}}\) \\ \relax 
		[nm] & [GWcm\textsuperscript{-2}] & [\textmu Jcm\textsuperscript{-2}] \\%
		\midrule
		1550 & \(\left(1.14\pm0.51\right)\times10^{0~}\) & \(\left(1.14\pm0.51\right)\times10^{2}\) \\
		2000 & \(\left(5.57\pm0.90\right)\times10^{-1}\) & \(\left(5.57\pm0.90\right)\times10^{1}\) \\
		2500 & \(\left(4.28\pm0.68\right)\times10^{-1}\) & \(\left(4.28\pm0.68\right)\times10^{1}\) \\
		2800 & \(\left(3.65\pm0.89\right)\times10^{-2}\) & \(\left(3.65\pm0.89\right)\times10^{0}\) \\
		3200 & \(\left(7.72\pm3.43\right)\times10^{-2}\) & \(\left(7.72\pm3.43\right)\times10^{0}\) \\
		3500 & \(\left(8.30\pm4.51\right)\times10^{-2}\) & \(\left(8.30\pm4.51\right)\times10^{0}\) \\
		\bottomrule
	\end{tabular}
\end{table}

\section{Beam profile}
\begin{figure}
	\FontSizeTable
\begin{subfigure}[]{8cm}%
	\includegraphics[width=\textwidth]{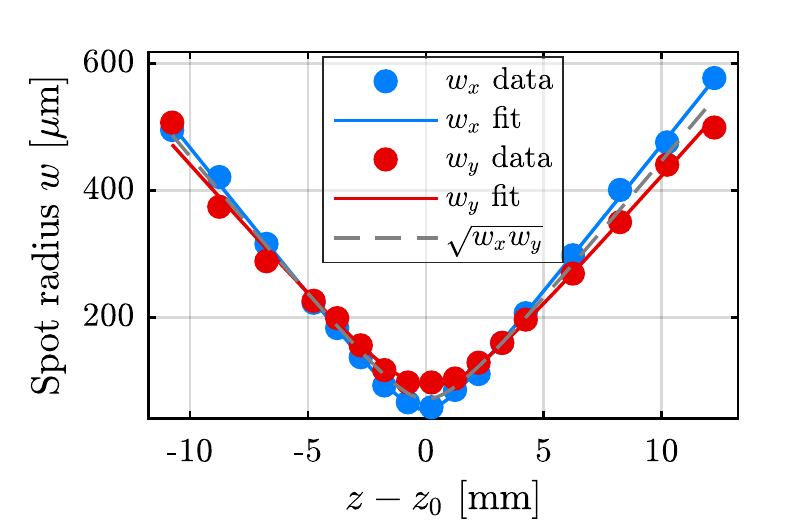}%
	\caption{Beam profile at 2800\,nm.}
	\label{Fig:2800M2}%
\end{subfigure}
\begin{subfigure}[]{8cm}%
	\includegraphics[width=\textwidth]{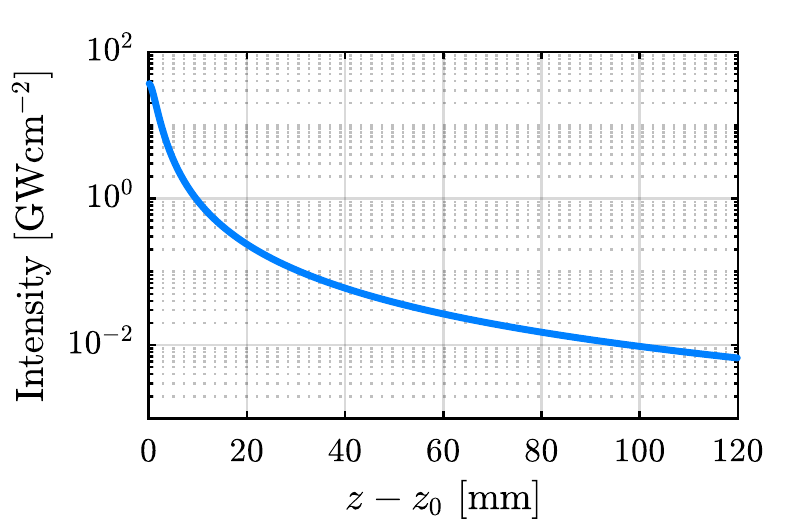}%
	\caption{Beam intensity at 2800\,nm.}
	\label{Fig:2800BeamIntensity}%
\end{subfigure}
\end{figure}

The spot size (radius) \( w(z) \) of the beam  was measured along the horizontal (\textit{x}-axis) and vertical (\textit{y}-axis) at 17 locations for each wavelength using a knife edge. The data was fitted to the spot size function for a Gaussian beam given by

\begin{equation}
w\left(z\right)=w_{0}\left[1+\left(\frac{M^{2}\lambda\left(z-z_{0}\right)}{\pi w_{0}^{2}}\right)^{2}\right]^{\frac{1}{2}}
\label{Eqn:SpotRadius}
\end{equation}

to determine the \( M^2 \) parameters and beam waist spot sizes \( w_0 \) of each beam. Then the spot size \( w(z) \) and intensity \( I(z) \) could be determined for any position \(  z  \) along the z-scan. The spot size measurement of the 2800\,nm beam is shown in Fig. \ref{Fig:2800M2} as an example. The intensity \( I(z) \) shown in Fig. \ref{Fig:2800BeamIntensity} spans four orders of magnitude along the z-scan range.

\section{Damage threshold}
A sacrificial sample of monolayer CVD graphene on \ce{CaF2} was used to find the  approximate damage threshold at each wavelength and provide a guide to an upper intensity limit for the z-scans. The laser induced damage was observed under an optical microscope (Olympus BX51) with 5\(\times\) and 10\(\times\) magnification. The maximum intensity used in each z-scan was less than one third of the observed damage threshold.

\begin{table}
	\caption{Damage thresholds of monolayer CVD graphene.}
	\label{Tbl:Ablation}
	\FontSizeTable
	\begin{tabular}{cccc}
	\toprule
	\(\lambda\) & \(w_0\) & \(I_{\mathrm{dam}}\) & \(F_{\mathrm{dam}}\) \\ \relax 
	[nm] & [\textmu m] & [GWcm\textsuperscript{-2}] & [mJcm\textsuperscript{-2}] \\%
	\midrule
	1550 & 68.8 & \(~59.8\pm10.9\) & \(2.99\pm0.54\) \\
	2000 & 58.5 & \(131.1\pm19.5\) & \(5.05\pm0.52\) \\
	2500 & 61.5 & \(191.2\pm32.7\) & \(9.56\pm1.64\) \\
	2800 & 59.1 & \(124.6\pm29.6\) & \(6.23\pm1.48\) \\
	3200 & 77.9 & \(116.6\pm~4.0\) & \(5.83\pm0.20\) \\
	3500 & 64.7 & \(144.9\pm21.5\) & \(7.25\pm1.08\) \\
	\bottomrule
	\end{tabular}
\end{table}

\section{FTIR spectra}
\begin{figure}
	\includegraphics[]{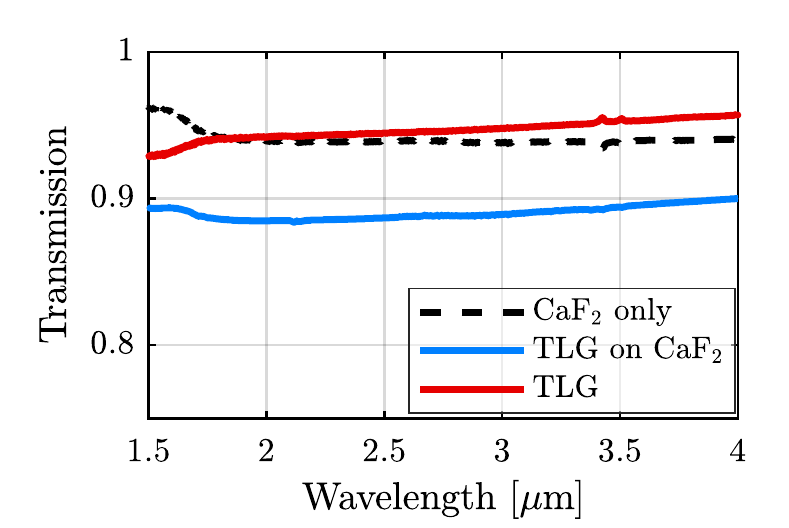}%
	\caption{FTIR spectra of a \ce{CaF2} substrate and the trilayer graphene (TLG) on \ce{CaF2}. The TLG transmission was calculated from the above mentioned spectra.}%
	\label{Fig:FTIR}
\end{figure}

Fourier transform infrared (FTIR) spectra of \ce{CaF2} with and without graphene are shown in Fig. \ref{Fig:FTIR}. The transmission increases a few percent with wavelength over the spectral region 1.5\,\textmu m to 4.0\,\textmu m, which is also observed in transmission measurements.

